\documentclass[12pt]{article}
\usepackage[utf8]{inputenc}
\usepackage{amsmath}
\usepackage{amssymb}
\usepackage{graphicx}

\makeatletter

\DeclareTextSymbolDefault{\textquotedbl}{T1}

\numberwithin{equation}{section}

\@ifundefined{date}{}{\date{}}

\usepackage{braket}
\usepackage{bm}
\usepackage{bbm}

\usepackage{epsfig}
\usepackage{pstricks}
\usepackage{color}
\usepackage{tikz}

\usepackage{hyperref}

\newcommand{\mq}{m}
\newcommand{\muq}{\mu}
\newcommand{\tmq}{\tilde{m}}
\newcommand{\tmuq}{\tilde{\mu}}

\@ifundefined{showcaptionsetup}{}{%
 \PassOptionsToPackage{caption=false}{subfig}}
\usepackage{subfig}
\makeatother

\begin{document}
\begin{titlepage}

\begin{flushright} 
KEK-TH-2230, RIKEN-QHP-479
\end{flushright} 

\vspace{1cm}

\begin{center}
{\bf \large 
Complex Langevin calculations in QCD at finite density}
\end{center}

\vspace{1cm}

\begin{center}
Yuta I{\sc to}$^{ab}$\footnote{E-mail address : y-itou@tokuyama.ac.jp},
Hideo M{\sc atsufuru}$^{c}$\footnote{E-mail address : hideo.matsufuru@kek.jp},
Yusuke N{\sc amekawa}$^{a}$\footnote{E-mail address : namekawa@post.kek.jp;
Present address is Yukawa Institute for Theoretical Physics, Kyoto University,  Kitashirakawa Oiwakecho, Sakyo-ku, Kyoto 606-8502 Japan},
Jun N{\sc ishimura}$^{ad}$\footnote{E-mail address : jnishi@post.kek.jp},\\
Shinji S{\sc himasaki}$^{e}$\footnote{E-mail address : shimasaki.s@gmail.com},
Asato T{\sc suchiya}$^{f}$\footnote{E-mail address : tsuchiya.asato@shizuoka.ac.jp},
Shoichiro T{\sc sutsui}$^{g}$\footnote{E-mail address : shoichiro.tsutsui@riken.jp}

\vspace{1cm}

$^a${\it Theory Center, Institute of Particle and Nuclear Studies,}\\
{\it High Energy Accelerator Research Organization (KEK),\\
1-1 Oho, Tsukuba, Ibaraki 305-0801, Japan} 

$^b${\it National Institute of Technology, Tokuyama College,\\
Gakuendai, Shunan, Yamaguchi 745-8585, Japan}

$^c${\it Computing Research Center, High Energy Accelerator 
Research Organization (KEK),\\
1-1 Oho, Tsukuba, Ibaraki 305-0801, Japan} 

$^d$\textit{Department of Particle and Nuclear Physics,}\\
\textit{School of High Energy Accelerator Science,}\\
{\it Graduate University for Advanced Studies (SOKENDAI),\\
1-1 Oho, Tsukuba, Ibaraki 305-0801, Japan} 

$^e${\it Research and Education Center for Natural Sciences, Keio University,\\
Hiyoshi 4-1-1, Yokohama, Kanagawa 223-8521, Japan} 

$^f${\it Department of Physics, Shizuoka University,\\
836 Ohya, Suruga-ku, Shizuoka 422-8529, Japan} 

$^g${\it Theoretical Research Division, Nishina Center, RIKEN,\\
Wako, Saitama 351-0198, Japan} 

\end{center}

\vspace{0.5cm}

\begin{abstract}
\noindent We demonstrate that
the complex Langevin method (CLM) enables calculations
in QCD at finite density in a parameter regime
in which conventional methods, such as 
the density of states method and the Taylor expansion method,  
are not applicable due to the severe sign problem.
Here we use the plaquette gauge action with $\beta = 5.7$ 
and 
four-flavor staggered fermions
with degenerate quark mass $\mq a = 0.01$ and 
nonzero quark chemical potential $\muq$.
We confirm that a sufficient condition for correct convergence
is satisfied 
for $\muq/T = 5.2 - 7.2$
on a $8^3 \times 16$ lattice
and 
$\muq/T = 1.6 - 9.6$ on a $16^3 \times 32$ lattice.
In particular, the expectation value of 
the quark number is found to have a plateau 
with respect to $\muq$
with the height of 24 for both lattices.
This plateau can be understood from the Fermi distribution of quarks,
and its height coincides with
the degrees of freedom of a single quark with zero momentum,
which is 3 (color) $\times$ 4 (flavor) $\times$
2 (spin) $=24$.
Our results may be viewed as the first step towards
the formation of the Fermi sphere, which plays a crucial role
in color superconductivity conjectured from effective theories.
\end{abstract}
\vfill
\end{titlepage}
\vfil\eject

\setcounter{footnote}{0}

\section{Introduction}

QCD at finite temperature and density attracts a lot of interest
due to its rich phase diagram predicted by effective theories.
Heavy-ion collision experiments are being performed
to elucidate the phase structure,
whereas the observation of gravitational waves is expected
to provide significant information on the equation of state of neutron stars
reflecting the phase structure.
In parallel, many efforts have been made toward
theoretical understanding of QCD at finite temperature and density.
The difficulty in theoretical analyses, however, is that 
nonperturbative calculations based on lattice QCD
suffer from the sign problem at finite density.
This problem occurs because 
of the complex fermion determinant,
which prevents us from applying standard Monte Carlo methods
based on important sampling by identifying the Boltzmann weight
as the probability distribution.

Here we focus on 
the complex Langevin 
method (CLM) \cite{Parisi:1984cs,Klauder:1983sp},
which has recently proven 
a promising method 
for solving the sign problem.
It is a complex extension of the stochastic quantization 
based on the Langevin equation,
where dynamical variables are complexified and 
physical quantities as well as the drift term 
are extended holomorphically.
An expectation value can be obtained as an average over the fictitious 
time evolution by the complex Langevin equation after thermalization.
See Ref.~\cite{Berger:2019odf} for a summary of the recent progress
concerning this method and other methods for solving the sign problem.
In particular, the CLM has been tested extensively in lattice QCD at finite 
density \cite{Sexty:2013ica,Aarts:2014bwa,Fodor:2015doa,Sinclair:2015kva,Sinclair:2016nbg,Sinclair:2017zhn,Sinclair:2018rbk,Nagata:2018mkb,Ito:2018jpo,Tsutsui:2018jva,Tsutsui:2019gwn,Kogut:2019qmi,Sexty:2019vqx,Sinclair:2019ysx,Tsutsui:2019suq,Scherzer:2020kiu}.

An important issue in the CLM is that
physical observables converge to wrong results depending on the parameters,
the model, or even on how the method is implemented.
It was not until recently that
the causes of this incorrect convergence 
were clarified
\cite{Aarts:2009uq,Aarts:2011ax,Nishimura:2015pba,Nagata:2015uga,Nagata:2016vkn,Salcedo:2016kyy,Aarts:2017vrv,Nagata:2018net}.
There are actually two kinds of causes;
one is the excursion problem \cite{Aarts:2009uq} and 
the other is the singular drift 
problem \cite{Mollgaard:2013qra,Nishimura:2015pba}.
In the case of finite density QCD, 
the excursion problem occurs 
when the link variables have long excursions away from 
the SU(3) group manifold.
This problem can be 
circumvented
by adding the gauge cooling procedure
after each Langevin update
as proposed in Refs.~\cite{Seiler:2012wz,Aarts:2013uxa} and 
justified in Refs.~\cite{Nagata:2015uga,Nagata:2016vkn}.
The singular drift problem occurs, on the other hand,
when
the fermion matrix has eigenvalues 
close to zero
since the drift term involves the inverse of the fermion matrix.

Both these problems can be detected
by just probing
the magnitude of the drift term \cite{Nagata:2016vkn},
which is calculated anyway for the Langevin evolution.
If the histogram of the drift
falls off exponentially or faster,
one can trust the results, 
as has been shown from a refined argument for 
justifying the CLM \cite{Nagata:2016vkn}
based on the discrete Langevin-time formulation.
The validity of this criterion
has been confirmed explicitly in
simple one-variable models \cite{Nagata:2016vkn}
as well as 
in exactly solvable models involving infinite degrees of freedom
such as chiral Random Matrix Theory \cite{Nagata:2018net}. 

An alternative criterion for correct convergence has been discussed 
from the viewpoint of 
the boundary terms \cite{Scherzer:2018hid,Scherzer:2019lrh},
which appear in the original argument \cite{Aarts:2009uq,Aarts:2011ax}
for justifying the CLM
based on the continuous Langevin-time formulation.
Note, however, that the limit of taking the Langevin-time stepsize to
zero and the notion of time-evolved observables,
which are crucial in the original argument,
can be subtle when the drift histogram does not fall off fast
enough \cite{Nagata:2016vkn}.
These subtleties are taken into account in the refined argument,
which led to the above criterion for the validity of the CLM.

In applications to QCD at finite density,
it is therefore of primary importance 
to determine the parameter region 
in which the CLM gives correct results. 
We address this issue
using staggered fermions
corresponding to QCD with four-flavor quarks,
which is known to have
a first order deconfining phase transition
at finite temperature $T$
for zero quark chemical potential $\muq=0$ \cite{Fukugita:1990vu}.
The phase structure
of 
four-flavor
QCD on the $T - \muq$ plane 
has been investigated by various methods including the CLM
using either
staggered fermions \cite{Sexty:2013ica,Fodor:2015doa,Nagata:2018mkb,Ito:2018jpo,Tsutsui:2018jva,Tsutsui:2019gwn,Tsutsui:2019suq,Fodor:2001au,DElia:2002tig,DElia:2004ani,Azcoiti:2004ri,Azcoiti:2005tv,Fodor:2007vv,DElia:2007bkz,Endrodi:2018zda} or Wilson fermions \cite{Scherzer:2020kiu,Chen:2004tb,deForcrand:2006ec,Li:2010qf,Takeda:2011vd,Jin:2013wta}.
The strong coupling expansion
was also applied
as reviewed in Refs.~\cite{Philipsen:2019rjq,Ohnishi:2015fhj}.

In our calculations, we use 
Wilson's plaquette action 
with 
$\beta = 5.7$
on $8^3 \times 16$ and $16^3 \times 32$ lattices.
The quark mass for the four-flavor staggered fermions
is set to $\mq a = 0.01$.
The criterion for correct convergence
of the CLM is found to be satisfied
for $\muq/T = 5.2 - 7.2$
on a $8^3 \times 16$ lattice
and 
$\muq/T = 1.6 - 9.6$ on a $16^3 \times 32$ lattice,
where conventional methods, 
such as the density of states method and the Taylor expansion method, 
are not applicable
due to the severe sign problem.

In particular, our calculations
reveal, for the first time, a plateau behavior of 
the quark number 
with respect to the quark chemical potential with the height of 24,
which can be understood from the Fermi distribution of quarks.
It actually coincides with 
the number of degrees of freedom for a single quark with zero momentum,
which is 3 (the number of colors) $\times$ 4 (the number of flavors)
$\times$ 2 (the number of spin degrees of freedom).
This can be regarded as the first step towards
the formation of the Fermi surface,
which plays a crucial role in color superconductivity. 
We also investigate other observables such as
the Polyakov loop and the chiral condensate.
Part of our results have been presented 
in proceedings 
articles \cite{Ito:2018jpo,Tsutsui:2018jva,Tsutsui:2019gwn,Tsutsui:2019suq}.

This paper is organized as follows.
In Section~\ref{sec:method} we explain how we apply the CLM
to lattice QCD at finite density.
In Section~\ref{sec:results} we discuss the validity of the CLM
in our simulation setup
and present our results with emphasis 
on their physical interpretation.
Section~\ref{sec:summary} is devoted to a summary and discussions.

\section{CLM for QCD at finite density}
\label{sec:method}

The partition function of QCD on a four-dimensional lattice
is given by
\begin{equation}
Z=
\int\prod_{x,\mu}dU_{x,\mu}\,\det M[U]\, e^{-S_{g}[U]}\ ,
\label{eq:partition-func}
\end{equation}
where $U_{x,\mu}\in$ SU(3) is a link variable
in the $\mu(=1,2,3,4)$ direction 
with $x=(x_{1},x_{2},x_{3},x_{4})$ representing a lattice site. 
For the gauge action $S_{g}[U]$, we use the plaquette action
\begin{equation}
S_{g}[U]=-\frac{\beta}{6}\sum_{x}\sum_{\mu\neq\nu}
\mathrm{tr}\left(U_{x,\mu}U_{x+\hat{\mu},\nu}U_{x+\hat{\nu},\mu}^{-1}U_{x,\nu}^{-1}
\right) \ ,
\label{eq:gauge-action}
\end{equation}
where $\beta = 6 / g^2$ with the gauge coupling $g$
and $\hat{\mu}$ represents the unit vector in the $\mu$ direction.
In this paper, we consider the 
four-flavor
staggered fermions
with degenerate quark mass $\mq$
and quark chemical potential $\muq$,
which corresponds to the fermion matrix
\begin{equation}
M_{xy}[U]=\tmq\delta_{xy}+
\sum_{\mu=1}^{4}\frac{\eta_{\mu}(x)}{2}
\left(e^{\tmuq\delta_{\mu4}}U_{x,\mu}
\delta_{x+\hat{\mu},y}-e^{-\tmuq\delta_{\mu4}}U_{x-\hat{\mu},\mu}^{-1}
\delta_{x-\hat{\mu},y}\right) \ ,
\label{eq:diracop}
\end{equation}
with
$\eta_{\mu}(x)\equiv(-1)^{x_{1}+\cdots+x_{\mu-1}}$ being
a site-dependent sign factor.
In \eqref{eq:diracop} we have defined
dimensionless parameters
$\tmq = \mq a$ and $\tmuq = \muq a$,
where $a$ represents the lattice spacing.
The determinant $\det M[U]$ in \eqref{eq:partition-func}
becomes complex 
for $\tmuq \neq 0$, which causes the sign problem.
Periodic boundary conditions are assumed except that
we impose anti-periodic boundary conditions
on the fermion fields in the temporal direction
so that the temporal extent of the lattice
represents the inverse temperature $T^{-1}$.

We apply the CLM to this system.
In this method, 
the link variables $U_{x,\mu}\in$ SU(3) are complexified 
to $\mathcal{U}_{x,\mu}\in$ SL(3,C),
and we consider a fictitious time evolution
governed by
\begin{equation}
\mathcal{U}_{x,\mu}(t+\varepsilon)=
\exp\left\{i\left(-\varepsilon v_{x,\mu}[\mathcal{U}(t)]+
\sqrt{\varepsilon}\eta_{x,\mu}(t)\right)\right\}\mathcal{U}_{x,\mu}(t) \ ,
\label{eq:Langevin-eq}
\end{equation}
which is the discrete version of the complex Langevin equation
with the stepsize $\varepsilon$.
Here $\eta_{x,\mu}(t)$ is the noise term, which is a traceless 
$3\times3$ Hermitian matrix
obeying the Gaussian distribution 
$\exp[-\frac{1}{4} \, \text{tr} \, \eta_{x,\mu}^{2}(t)]$
so that
\begin{equation}
\langle\eta_{x,\mu}^{ij}(s)\eta_{y,\nu}^{kl}(t)\rangle_{\eta}
=2\delta_{x y}\delta_{\mu\nu}\delta_{st} 
\left( \delta_{il}\delta_{jk}- \frac{1}{3}\delta_{ij}\delta_{kl} \right)\ ,
\end{equation}
where the symbol $\langle \ \cdot \ \rangle_{\eta}$ represents 
the expectation value with respect to the Gaussian noise $\eta$.
The $v_{x,\mu}$ in eq.~(\ref{eq:Langevin-eq}) is 
the drift term, which is defined by holomorphically extending
\begin{equation}
v_{x,\mu}[U]=\left.\sum_{a}\lambda_{a}\frac{\partial}{\partial\alpha}
S[e^{i\alpha\lambda_{a}}U_{x,\mu}]\right|_{\alpha=0} 
\label{eq:drift-term}
\end{equation}
defined for 
unitary configurations $U$,
where $S[U]=S_{g}[U]-\ln\det M[U]$
and $\lambda_{a}\;(a=1,\ldots, 8)$ are the SU(3) generators
normalized by 
$\text{tr}(\lambda_{a}\lambda_{b})=\delta_{ab}$.
Note that the drift term is not Hermitian for $\tmuq \neq 0$
even for unitary link variables,
which makes the time-evolved link variables inevitably non-unitary.

In order to calculate the drift term
\begin{align}
v_{x , \mu}^{\rm (f)}   &=
- \sum_{a}\lambda_{a} \mathrm{tr} \, \left( M^{-1} 
\left.\frac{\partial}{\partial\alpha} 
M[e^{i\alpha\lambda_{a}}U_{x,\mu}]\right|_{\alpha=0} \right) 
%
  \label{fermion-drift}
\end{align}
obtained from the fermion determinant,
we use the standard noisy estimator.
See Appendix A of Ref.~\cite{Nagata:2018mkb} for the details.

The expectation value of an observable 
$\mathcal{O}(\mathcal{U})$ is calculated as
\begin{equation}
\bar{\mathcal{O}} =
\frac{1}{\tau}\int_{t_{0}}^{t_{0}+\tau} dt\,\mathcal{O}(\mathcal{U}(t)) 
\label{eq:vev}
\end{equation}
using the notation of continuum Langevin time $t$ for simplicity.
Here $t_0$ represents the time needed for thermalization
and $\tau$ represents the total Langevin time for taking the average, 
which should be large enough to achieve good statistics.
The CLM is justified
if the expectation value $\bar{\mathcal{O}}$ 
obtained by the CLM agrees with the expectation value
\begin{equation}
\langle\mathcal{O}(U)\rangle = \frac{1}{Z}\int dU\,
\mathcal{O}(U)\det M[U] \, e^{-S_{g}[U]} 
\end{equation}
defined in the path integral formalism.

The criterion for justification \cite{Nagata:2016vkn}
is based
on the magnitude of the drift term (\ref{eq:drift-term})
\begin{equation}
v=\underset{x,\mu}{\max}
\sqrt{\frac{1}{3}
{\rm tr} \, \Big( v_{x,\mu}^\dagger v_{x,\mu} \Big) }
\ .
\label{eq:mag_drift}
\end{equation}
If the histogram of this quantity 
falls off exponentially or faster,
we can trust the results.
This can be violated either by the excursion problem 
or by the singular drift problem as we mentioned in the Introduction.
In the former case, it is the drift coming from the gauge action
that shows slow fall-off in the histogram,
while in the latter case, it is the drift coming from the 
fermion determinant.

The excursion problem can also be probed
by the unitarity norm
\begin{equation}
\mathcal{N}
=\frac{1}{12 L_{\text{t}} L_{\text{s}}^{3}}
\sum_{x,\mu}\mathrm{tr}\, (\mathcal{U}_{x,\mu}^{\dagger}
\mathcal{U}_{x,\mu}-\mathbf{1}) \ ,
\label{eq:unitary-norm}
\end{equation}
which represents the distance of a configuration
from the SU(3) manifold
with $L_{\text{t}}$ and $L_{\text{s}}$ being the lattice size 
in the temporal and spatial directions, respectively.
The unitarity norm (\ref{eq:unitary-norm})
is positive semi-definite and it becomes zero if and only 
if all the link variables are unitary.
Rapid growth of the unitarity norm typically signals
the occurrence of the excursion problem.
We note, however,
that recent work \cite{Hirasawa:2020bnl}
on 2D U(1) theory with a $\theta$ 
term suggests that the CLM works
even if the unitarity norm becomes large 
as far as the drift histogram falls off fast.\footnote{The fact that
a large unitarity norm does not necessarily mean incorrect results
has been noticed for a long time; See, for instance, 
Ref.~\cite{Berges:2007nr}.}
Therefore, one cannot tell the validity of the CLM
by looking at the unitarity norm alone.
 
In order to avoid the excursion problem,
we use the gauge cooling \cite{Seiler:2012wz,Aarts:2013uxa},
which amounts to making a complexified gauge transformation
\begin{equation}
\delta_g \, \mathcal{U}_{x,\mu} = g_x \mathcal{U}_{x,\mu} g_{x + \hat{\mu}}^{-1}
\ , \quad 
g\in\text{SL(3,C)} 
\label{eq:gauge-tf}
\end{equation}
in such a way that
the unitarity norm is minimized.
Adding this procedure after each step of the Langevin-time evolution
does not spoil the argument for justification as is shown
in Refs.~\cite{Nagata:2015uga,Nagata:2016vkn}.
The gauge cooling keeps all the link variables as close to unitary
matrices as possible during the Langevin-time evolution.

As for physical observables,
we calculate the Polyakov loop, the quark number
and the chiral condensate.
The Polyakov loop is given by
\begin{equation}
 P=\frac{1}{3L_{\mathrm{s}}^{3}} \sum_{\vec{x}} \mathrm{tr} 
\left( \prod_{x_4=1}^{L_{\rm t}} U_{(\vec{x},x_4),4} 
\right)
\ , \label{eq:polyakov}
\end{equation}
with $\vec{x}=(x_{1},x_{2},x_{3})$ being a spatial coordinate on the lattice.
The quark number 
$N_{\rm q}$ and the chiral condensate $\Sigma$ are defined by 
\begin{eqnarray}
 N_{\rm q} 
 &=& \frac{1}{L_{\rm t}} \frac{\partial}{\partial\tmuq}\log Z
 = \frac{1}{L_{\rm t}}
\left\langle 
\sum_{x}\frac{\eta_{4}(x)}{2}\mathrm{tr}
\left(e^{\tmuq}M_{x+\hat{4},x}^{-1}U_{x,4}
+e^{-\tmuq}M_{x-\hat{4},x}^{-1}U_{x-\hat{4},4}^{-1}\right)\right\rangle \ , 
\label{eq:quark-number}
 \\
\Sigma 
 &=& \frac{1}{L_{\rm s}^3 L_{\rm t}}
\frac{\partial}{\partial \tmq}\log Z
 = \left\langle \mathrm{Tr} \, M^{-1} \right\rangle \ ,
\label{eq:chiral-condensate}
\end{eqnarray}
where the latter trace ${\rm Tr}$ is taken not only for the color index 
but also for
the spacetime index. 
These two quantities (\ref{eq:quark-number}) and (\ref{eq:chiral-condensate}) are calculated 
by the so-called noisy estimator using 20 noise vectors.

\section{Results}
\label{sec:results}

We have performed 
complex Langevin
simulations at $\beta=5.7$ 
with degenerate quark mass
$\tmq=0.01$
on $8^{3}\times16$ and $16^{3}\times32$ lattices.
We determine the lattice spacing as $a^{-1} = 4.65(1)$ GeV
from the Sommer scale \cite{Sommer:1993ce} 
by performing independent 
Hybrid Monte Carlo simulations
on a $24^{3}\times48$ lattice with $\tmuq = 0$.
The quark chemical potential is varied within the range
$\tmuq = 0.1 - 0.5$ 
on the $8^{3}\times16$ lattice
and $\tmuq = 0.05 - 0.325$ 
on the $16^{3}\times32$ lattice.
The Langevin-time stepsize is chosen initially as 
$\varepsilon=10^{-4}$ and reduced adaptively \cite{Aarts:2009dg}
when the magnitude (\ref{eq:mag_drift}) 
of the drift becomes larger than 
the threshold 100.
Measurement of the observables is made every $10^{-2}$ Langevin time. 
The total Langevin time after thermalization
is $\tau =70-140$ 
for the $8^{3}\times16$ lattice
and $\tau =10-20$ for the $16^{3}\times32$ lattice.
The error bars of the physical observables are evaluated 
by the jackknife method with the bin sizes of $2$ Langevin time 
on $8^3 \times 16$ and $0.5$ Langevin time on $16^3 \times 32$.

\subsection{Validity of the CLM}
\label{sec:validity}

\begin{figure}
\centering{}
\includegraphics{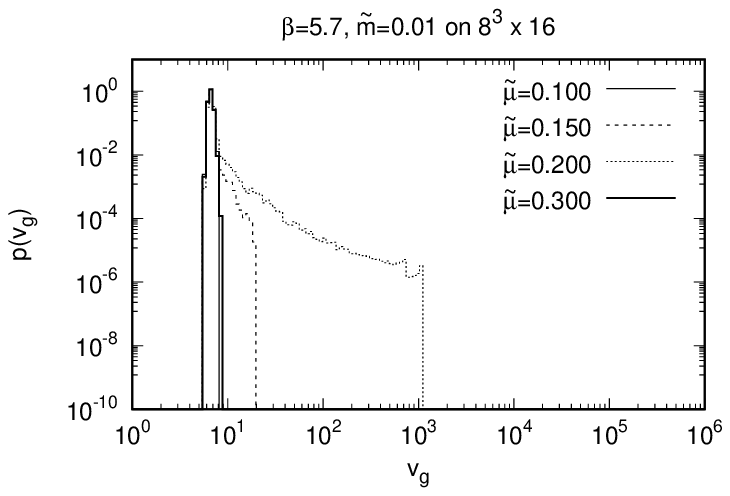}
\includegraphics{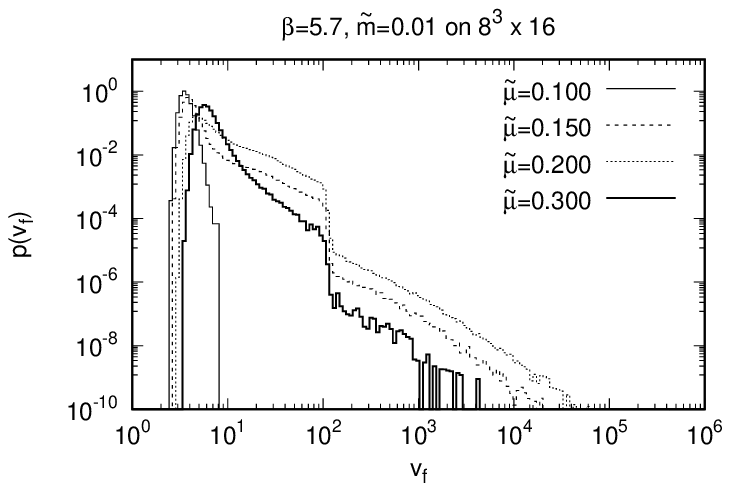}

\includegraphics{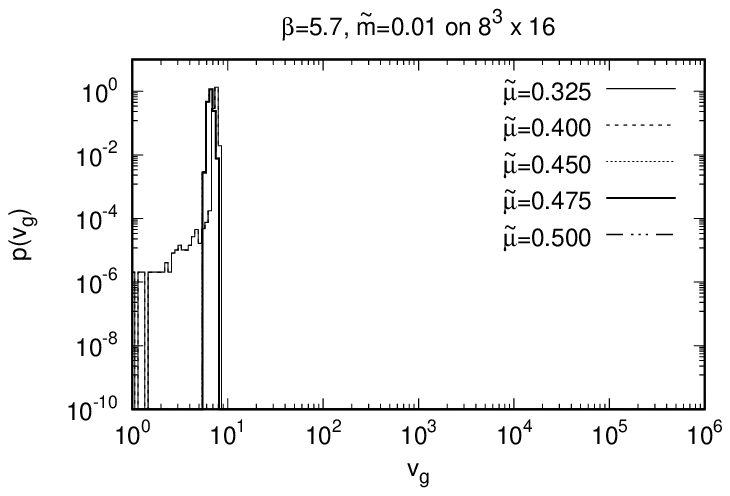}
\includegraphics{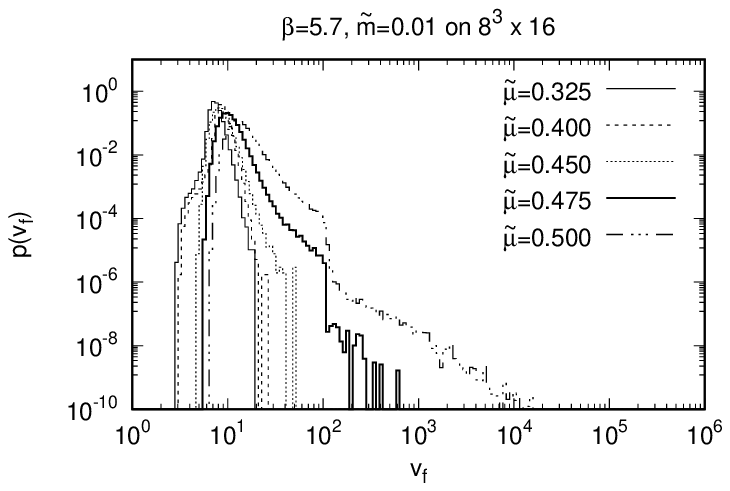}
\caption{The probability distributions $p(v)$ of the drift term 
obtained from simulations on a $8^{3}\times16$ lattice
with $\beta=5.7$ and $\tmq=0.01$
are plotted for the drift term $v_{\rm g}$ 
coming from the gauge action (Left)
and $v_{\rm f}$ 
from the fermion determinant (Right).
The upper and lower panels show
the data points for $\tmuq\leq 0.3$ and $\tmuq\geq 0.325$, 
respectively.}
\label{fig:drift-8x16-q001}
\end{figure}

\begin{figure}
\centering{}
\includegraphics{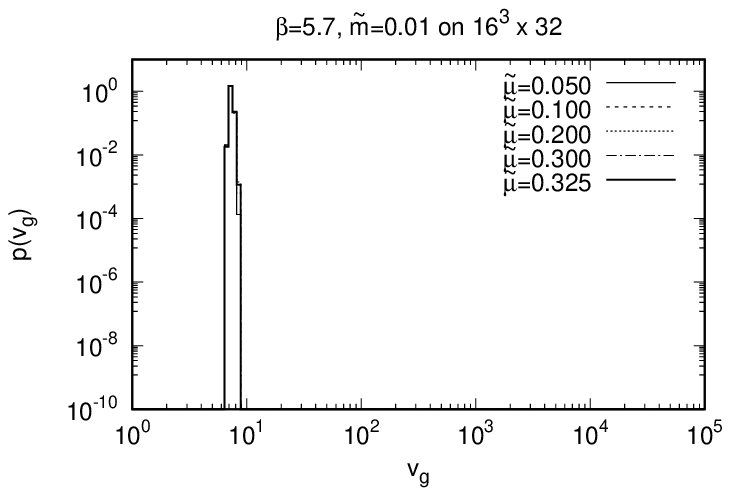}
\includegraphics{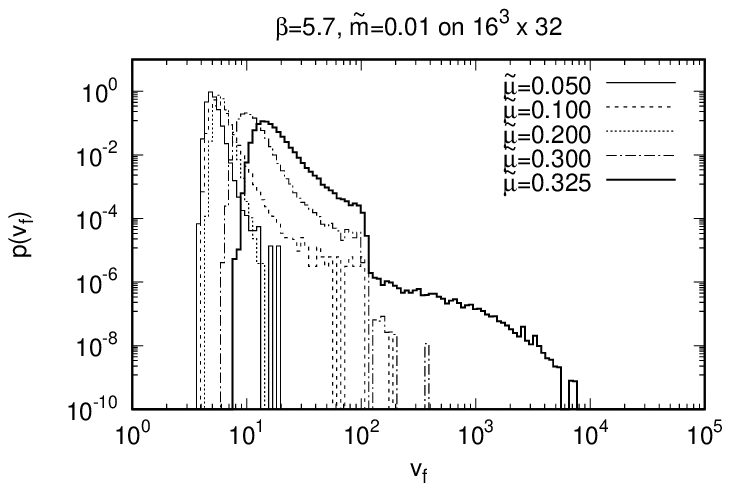}
\caption{The probability distributions $p(v)$ of the drift term 
obtained from simulations on a $16^{3}\times32$ lattice
with $\beta=5.7$ and $\tmq=0.01$
are plotted
for the drift term $v_{\rm g}$ coming from the gauge action (Left)
and $v_{\rm f}$
from the fermion determinant (Right).}
\label{fig:drift-16}
\end{figure}

Let us first discuss
the validity
of the CLM 
in our simulation setup.
In Fig.~\ref{fig:drift-8x16-q001}
we plot the distribution $p(v)$ for
the magnitude of the drift term (\ref{eq:mag_drift}) coming from the 
gauge action (Left) 
and from the fermion determinant (Right), respectively,
obtained by simulations on a $8^{3}\times16$ lattice 
with $\beta=5.7$ and $\tmq=0.01$.
For the sake of visibility, we separate the data points
into two regions $\tmuq\leq 0.3$ and $\tmuq\geq 0.325$ 
and show them in the upper and lower panels, respectively.
We find that 
the drift term coming from the gauge action 
shows slow fall-off for $\tmuq= 0.2$,
which implies that the excursion problem occurs only in this case.
The drift term coming from the fermion determinant, on the other hand,
shows slow fall-off for
$0.15\leq\tmuq\leq0.3$ and $0.475\leq\tmuq$
indicating the occurrence of the singular drift problem 
in these cases.\footnote{The sudden drop of the histogram
around $v_{\rm f}=100$, which is also seen in 
Fig.~\ref{fig:drift-16} (Right),
is an artifact of the chosen threshold for the 
adaptive stepsize mentioned at the beginning of 
Section \ref{sec:results}.}
Thus we conclude 
that the CLM gives correct results in the regions
$\tmuq \simeq 0.1$ and $0.325\leq\tmuq\leq0.45$
on the $8^{3}\times16$ lattice.

\begin{figure}
\includegraphics{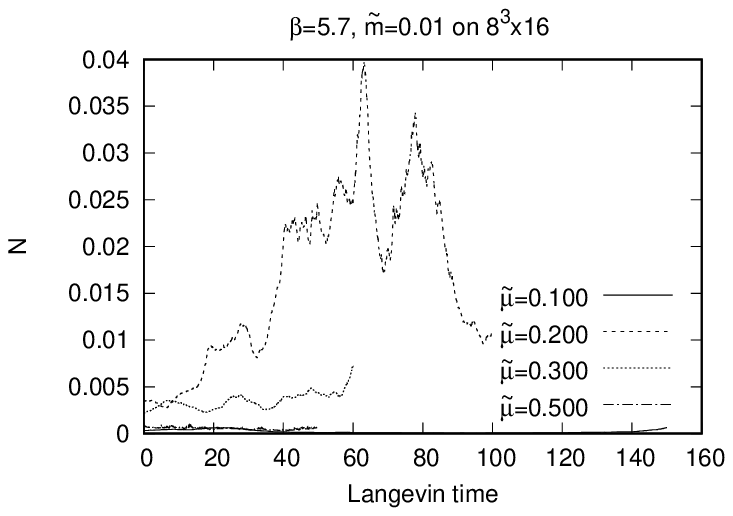}
\includegraphics{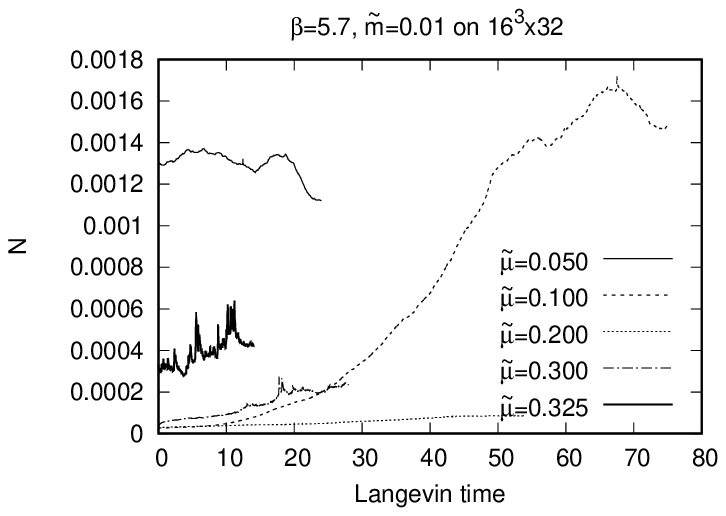}
\centering{}\caption{The Langevin-time histories of the 
unitarity norm (starting from the initial configuration)
are plotted for $\beta=5.7$ and $\tmq=0.01$ on the 
$8^{3}\times16$ lattice (Left) and the $16^{3}\times32$ lattice (Right).}
\label{fig:unitary-norm}
\end{figure}

Figure \ref{fig:drift-16} shows similar plots
for a $16^{3}\times32$ lattice with the same 
$\beta=5.7$ and $\tmq=0.01$.
Here we find that the excursion problem does not occur for any values 
of $\tmuq$, whereas the singular drift problem
occurs for $\tmuq=0.325$.
Thus we find that the CLM is expected to give correct results
for $0.05 \leq \tmuq \leq 0.3$
on the $16^{3}\times32$ lattice.

In Fig.~\ref{fig:unitary-norm}, we plot the Langevin-time histories 
of the unitarity norm (\ref{eq:unitary-norm}).
For the $8^{3}\times16$ lattice (Left),
we find that the history for $\tmuq= 0.2$ looks quite violent,
which is consistent with what we observe 
in Fig.~\ref{fig:drift-8x16-q001} (Top-Left).
For the $16^{3}\times 32$ lattice (Right), on the other hand,
the unitarity norm seems to be well under control;
Note that the scale of the vertical axis here
is an order of magnitude smaller than that in the Left panel.
This is also consistent with what we observe in
Fig.~\ref{fig:drift-16} (Left).
Let us emphasize, however, that from the histories of the unitarity 
norm alone, we cannot judge the validity of the CLM unambiguously.
Note also that
the unitarity norm has a long autocorrelation time
as one can see from Fig.~\ref{fig:unitary-norm}.
Fortunately, we find that 
the physical observables we investigate
are not correlated with the unitarity norm,
and they have a much shorter autocorrelation time.
This is important because it allows us 
to calculate their expectation values
reliably within a reasonable length of the total Langevin time.

\subsection{Physical observables}
\label{sec:observables}

\begin{figure}
\centering{}
\includegraphics[width=9cm]{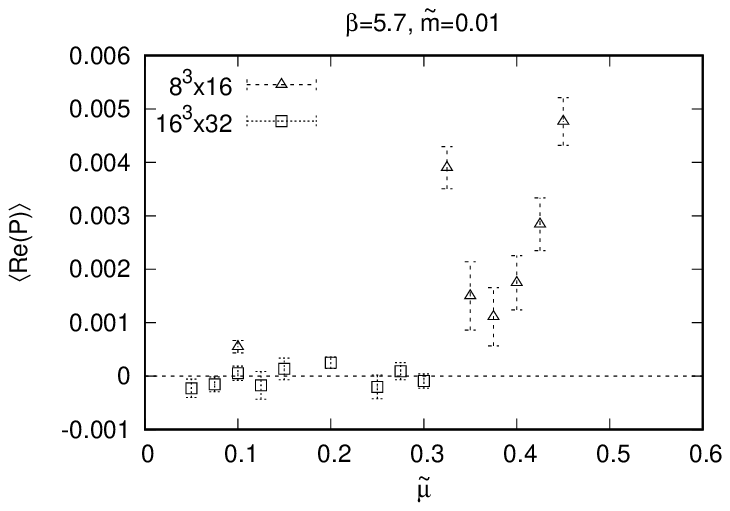}
\includegraphics[width=9cm]{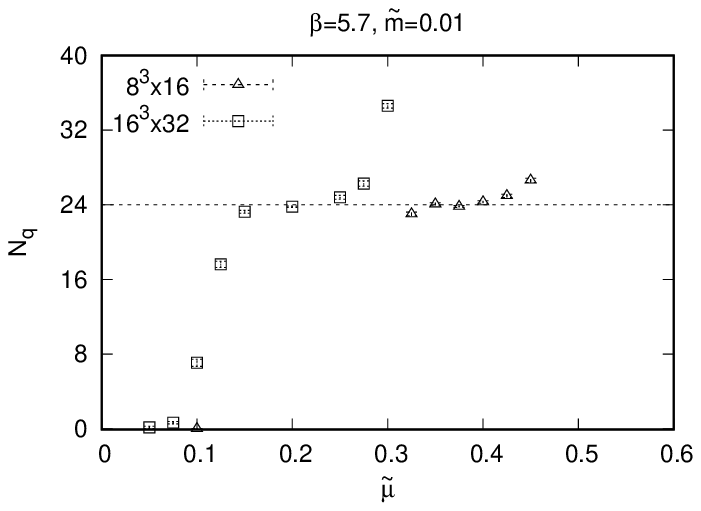}
\includegraphics[width=9cm]{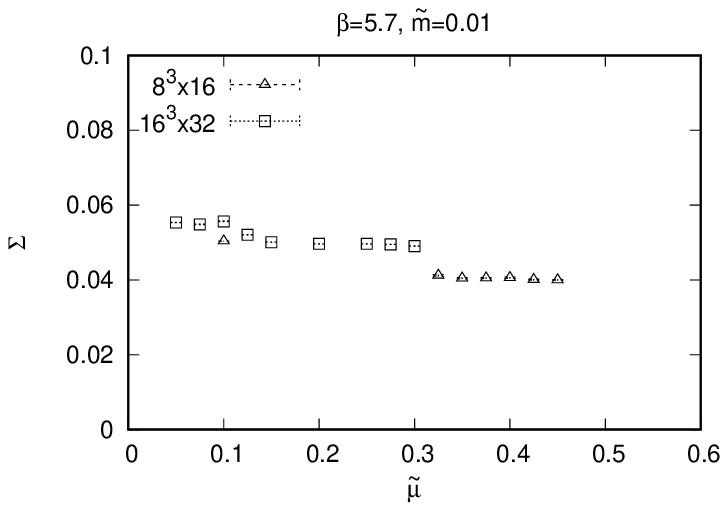}
\caption{The real part of the Polyakov loop $P$ (Top),
the quark number $N_{\rm q}$ (Middle)
and the chiral condensate $\Sigma$ (Bottom)
are plotted against the quark chemical potential $\tmuq$ 
for $\beta=5.7$ and $\tmq=0.01$
on the $8^3 \times 16$ and $16^{3}\times32$ lattices.
}
\label{fig:polyakov}
\end{figure}

In what follows, we present only the data points
in the parameter region in which 
the criterion for justification is satisfied.
In Fig.~\ref{fig:polyakov} (Top) we plot
the real part of 
the Polyakov loop (\ref{eq:polyakov})
against 
$\tmuq$
for 
$\beta=5.7$ and $\tmq=0.01$ on the $8^{3}\times16$ and 
the $16^{3}\times32$ lattices.
The results 
for the $8^3 \times 16$ lattice are slightly nonzero
and the results for the $16^{3}\times32$ lattice are
consistent with zero.
Let us recall here that the Polyakov loop
is an order parameter for the deconfining transition.
The interpretation of our results requires some care, though.
Note that
the spatial size of our lattice is $aL_{\text{s}}=0.36{\rm ~fm}$ 
and $0.68{\rm ~fm}$ for $L_{\text{s}}=8$ and 16, respectively,
which are smaller than 
the typical length scale of QCD, namely $\Lambda_{\rm LQCD}^{-1} \sim  1{\rm ~fm}$.
Thus the situation we are simulating should be regarded as
QCD in a small box, where the notion of quark confinement 
does not make sense.
In fact, the temperature is $T\sim290{\rm ~MeV}$ and $145{\rm ~MeV}$ 
for $L_{\text{t}}=16$ and 32, respectively, which are higher than or 
close to the critical temperature $T_{\text{c}} \sim 170$ MeV 
for the deconfining transition in 
QCD with 
four-flavor staggered quarks \cite{Engels:1996ag}
with $m / T = 0.2$, where large physical volume is implicitly assumed.
The Polyakov loop being either small or zero in our setup
simply confirms that we are probing the ``low temperature'' behavior
of such a finite size system due to the chosen aspect ratio of our lattice.

In Fig.~\ref{fig:polyakov} (Middle) we plot
the quark number (\ref{eq:quark-number})
against $\tmuq$
for $\beta=5.7$ and $\tmq=0.01$ on 
the $8^{3}\times16$ and $16^{3}\times32$ lattices.
On both lattices, we observe a plateau at 
the height of $N_{\rm q}=24$.
In order to understand this behavior,
let us recall that 
the physical extent of our lattice
is too small to create a baryon in it.
The effective gauge coupling is small due to the
asymptotic freedom, which makes 
the Fermi distribution of quarks
qualitatively valid.
At sufficiently large $\tmuq$,
the path integral is therefore dominated 
by a state obtained from the vacuum by creating quarks
with momentum $\vec{p}$ satisfying
$\sqrt{\vec{p}^{2}+m_{\rm eff}^{2}} \leq \tmuq$,
where $m_{\rm eff}$ is the effective mass including quantum 
corrections\footnote{Here 
we neglect finite lattice spacing effects 
assuming that we are close to the continuum limit.}.
It should be noted here that the momentum is discretized 
in a finite box as $\vec{p}=(2\pi/L_{\text{s}})\, \vec{n}$ 
with $\vec{n}$ being a 3D integer vector.
In particular, for $m_{\rm eff} \leq \tmuq \leq \mu_1$, 
where $\mu_1=\sqrt{(2\pi/L_{\text{s}})^2 +m_{\rm eff}^{2}}$, 
only the quarks with $\vec{p}=0$ are created.
The height of the plateau is therefore given
by the internal degrees of freedom
$N_{\text{f}} \times N_{\text{c}} \times N_{\text{spin}} = 4 \times 3 \times 2 =24$,
where $N_{\text{f}}$, $N_{\text{c}}$ and $N_{\text{spin}}$
are the number of flavors, the number of colors and 
the number of spin degrees of freedom, respectively\footnote{In the 
case of two-flavor Wilson fermions, the height of the plateau
is found to be $12$, which agrees with
$N_{\text{f}} \times N_{\text{c}} \times N_{\text{spin}} = 2 \times 3 \times 2 =12$.
We thank Manuel Scherzer for confirming this with his 
data \cite{Scherzer:2019weu}.}.
In Fig.~\ref{fig:polyakov} (Middle), we also
observe that our data start to leave the plateau for larger $\tmuq$,
which can be understood as the effects of quarks 
with the first non-zero momenta being created.
The value of $\tmuq$ at which this growth of $N_{\rm q}$ occurs
is smaller than $\mu_1$ defined above,
which can be understood as a result of 
finite temperature effects.\footnote{For instance, 
for the $16^{3}\times32$ lattice, 
a sharp transition to the second plateau with the height 
$N_{\rm q}=24\times(1+6)=168$ is expected to occur
at $\mu_1 \sim 0.4$ using $m_{\rm eff}^{2}\sim 0.1$
if the temperature were zero. However, our results
show an increase of $N_{\rm q}$ already at $\tmuq \sim 0.3$
due to finite temperature effects.}
Note that $\mu_1$ becomes smaller as the lattice becomes larger,
which is clearly reflected in our results.

The appearance of such plateaus in the quark number
for QCD in a finite box
was discussed in the case of free theory 
using the naive lattice action for fermions \cite{Matsuoka:1983wy}.
See also Ref.~\cite{Hands:2010zp} for such behaviors
based on one-loop perturbative calculations
in the continuum QCD on a small ${\rm S}^3$.
In a separate paper \cite{Yokota:2020aaa},
we will report on our results of the CLM for larger $\beta$,
which are compared with perturbative results 
obtained with staggered fermions.
This plateau behavior should not be confused with
the quark number saturation that occurs at much larger $\tmuq$.
In that case, the path integral is dominated
by a state with all the sites being occupied by fermions.
Taking the internal degrees of freedom into account,
the height of the plateau becomes
$N_{\text{f}}\times N_{\text{c}}\times N_{\text{spin}} \times L_{\rm s}^3
= 24 \times L_{\rm s}^3$, which is much higher than $24$.

In Fig.~\ref{fig:polyakov} (Bottom)
we show our results for the chiral condensate (\ref{eq:chiral-condensate}).
The plateau behaviors appear here as well
because of the ``low temperature'', where
changing $\tmuq$ a little cannot create quarks
at higher energy levels.
The plateau corresponding to
the state with the zero-momentum
quarks
appears 
with the height only slightly lower than that corresponding to
the state that dominates at $\tmuq=0$.
This suggests that the chiral symmetry for $\tmq=0$,
which is considered to be spontaneously broken at $\tmuq=0$,
does not get restored by increasing $\tmuq$
in the present parameter regime.

The plateau behaviors observed above 
in the quark number and the chiral 
condensate are analogous to those in
two-color QCD\footnote{This model has no sign problem even
at finite density and hence the standard hybrid Monte Carlo algorithm
is applicable.}
using two-flavor Wilson fermions on a $3^3 \times 64$ lattice 
at $\beta=24$ with finite $\mu$ \cite{Hands:2010vw}.
In that case, however, the height of the plateau 
in the quark number does not agree with the 
free fermion results, which is in contrast to
our results for the SU(3) gauge group on $8^3 \times 16$
and $16^3 \times 32$ lattices.

\section{Summary and Discussions}
\label{sec:summary}

In this paper we have applied the CLM to QCD at finite density
with the plaquette gauge action and the four-flavor staggered fermions
on $8^3 \times 16$ and $16^3 \times 32$ lattices.
While the spatial size of our lattice is still as small as
$aL_{\text{s}}=0.36{\rm ~fm}$ 
and $0.68{\rm ~fm}$ for $L_{\text{s}}=8$ and 16,
we find that the criterion for correct convergence
is satisfied
for $\muq = 1.5-2.1\mbox{~GeV}$ 
on the $8^3 \times 16$ lattice
and for $\muq = 0.23-1.4\mbox{~GeV}$ 
on the $16^3 \times 32$ lattice
with $T \sim 290\mbox{~MeV}$ and $145\mbox{~MeV}$, respectively.
These parameter regimes
cannot be reached by conventional methods, 
such as the density of states method and the Taylor expansion method.
Thus our results clearly demonstrate a big advantage
of the CLM in overcoming the sign problem in finite density QCD.

Let us also mention that the previous work \cite{Nagata:2018mkb}
shows that
the CLM works on a $4^3 \times 8$ lattice with the same $\beta$ 
but only with the aid of the deformation technique \cite{Ito:2016efb}, 
which is actually not needed for the lattice size in the present work.
Thus we find that the situation becomes better for a larger lattice,
which is also seen 
by comparing our results for 
$8^3 \times 16$ and $16^3 \times 32$ lattices 
in section \ref{sec:validity}.

One of our main physical results is that
the quark number exhibits a plateau behavior
as a function of the quark chemical potential with the height of 24
at sufficiently large $\muq$.
This has been interpreted as the creation of
quarks with zero momentum, 
which has the internal degrees of freedom 
$N_{\text{f}} \times N_{\text{c}} \times N_{\text{spin}} = 4 \times 3 \times 2 =24$.
We may regard it as the first step towards
the formation of the Fermi surface, 
which plays a crucial role in color superconductivity.

It is of particular importance to perform similar calculations
with larger lattices.
That will 
enable us to observe 
the growth of the Fermi sphere with moderate values of $\muq$
thanks to better momentum resolution.
Note that color superconductivity is expected to occur
due to Cooper pairing of quarks near the Fermi surface, 
which is possible even at weak coupling or 
in a small physical volume \cite{Amore:2001uf}.
For this reason,
we are currently exploring the larger $\beta$ regime,
where we can compare our results against
perturbative calculations \cite{Yokota:2020aaa}.
In particular, we are trying to observe
a departure from perturbative behaviors as $\beta$ gets
smaller than some critical value, which would signal the onset of 
color superconductivity.

\section*{Acknowledgements}

We thank E.\ Itou, A.\ Ohnishi and M.\ Scherzer for important discussions,
which enabled us to establish the interpretation of the plateau behavior 
observed in our results for the quark number.
The authors are also grateful to
M.P.\ Lombardo for bringing our attention to Ref.~\cite{Matsuoka:1983wy}
and to Y.\ Asano, T.\ Kaneko and T.\ Yokota 
for collaborations in the ongoing projects related to the present work.
This research was supported by MEXT as 
``Priority Issue on Post-K computer''
(Elucidation of the Fundamental Laws and Evolution of the Universe) 
and as 
``Program for Promoting Researches on the Supercomputer 
Fugaku''
(Simulation for basic science: 
from fundamental laws of particles to creation of nuclei).
It is also supported by Joint Institute for 
Computational Fundamental Science (JICFuS).
Computations were carried out using computational resources of the K computer
provided by the RIKEN Center for Computational Science
through the HPCI System Research project (Project ID:hp180178, hp190159) 
and the Oakbridge-CX provided by the Information Technology Center at 
the University of Tokyo through the HPCI System Research project (Project ID:hp200079).
Y.\ N.\ and J.\ N.\ were supported in part by JSPS KAKENHI Grant 
Numbers JP16H03988, JP18K03638.
S.\ S.\ was supported by the MEXT-Supported Program for the Strategic 
Research Foundation at Private 
Universities ``Topological Science'' (Grant No. S1511006).
S.\ T.\ was supported by the RIKEN Special Postdoctoral Researchers Program.

\bibliographystyle{JHEP}
\bibliography{qcd-clm}

\end{document}